\documentclass[fleqn,twoside]{article}
\usepackage[headings]{espcrc2}

\readRCS
$Id: espcrc2.tex,v 1.2 2004/02/24 11:22:11 spepping Exp $
\usepackage{amssymb}
\usepackage{amsfonts}
\usepackage{amsmath}
\usepackage{mathrsfs}
\usepackage{eufrak}
\usepackage{fancyhdr}
\usepackage{verbatim}
\usepackage[dvips]{graphicx}
\DeclareMathAlphabet{\mathpzc}{OT1}{pzc}{m}{it}
\newcommand{\sectioncount}[1]{\section{#1}\setcounter{equation}{0}}

\newcommand{\oneover}[1]{\frac{1}{#1}}

\newcommand{\rhofo}{\varrho}

\newcommand{\cqq}{\check q}

\newcommand{\crho}{\check\rho}

\newcommand{\crhofo}{\check\varrho}

\newcommand{\weq}{w_{\scriptscriptstyle{(eq)}}}
\newcommand{\Mpl}{M_{Pl}^4}

\newcommand{\diag}{\mathrm{diag}}

\newcommand{\bea}{\begin{eqnarray}}
\newcommand{\ena}{\end{eqnarray}}
\newcommand {\non}{\nonumber}

\newcommand{\refeq}[1]{(\ref{#1})}

\newcommand{\g}{\gamma}

\newcommand{\z}{\zeta}

\renewcommand{\k}{\kappa}

\renewcommand{\L}{\Lambda}
\newcommand{\m}{\mu}

\newcommand{\n}{\nu}

\newcommand{\x}{\xi}

\newcommand{\p}{\pi}

\newcommand{\s}{\sigma}

\renewcommand{\AmS}{{\protect\the\textfont2
  A\kern-.1667em\lower.5ex\hbox{M}\kern-.125emS}}
\title{Holography in 7D Randall--Sundrum cosmology}
\author{
Liuba Mazzanti%
\address{Milano--Bicocca University and INFN Milano--Bicocca, 20126 Milano, ITALY%
\\~ and CPhT, \'Ecole Polytechnique, 91128 Palaiseau, FRANCE}%
}
\runauthor{L. Mazzanti}
\begin{document}
\begin{abstract}
This work is based on \cite{Mazzanti:2007sg}.
We define the 7D Randall--Sundrum (RS) background and construct the corresponding 6D theory via AdS/CFT duality. In simple approximations, the Friedmann equation derived from 7D RS cosmology and the brane
view--point results exactly match. Possible cosmologies on the bulk side are also outlined. 
\end{abstract}
\maketitle
\sectioncount{Introduction and set--up}\label{introduction}
Brane--worlds are insightful models for both Standard Model building and cosmology. Matter fields can be localized on the (eventually warped) brane while gravitons are in principle free to propagate in all
extended dimensions. In the 5D RS model \cite{Randall:1999ee} gravity is effectively 4D in the IR, due to the RS brane cutting the space--time. We consider a 7D RS set--up, with a 5--brane warped over
a compact 2D space, a $\mathbb{Z}_2$ reflection of the seventh direction (tranverse to the brane) $z\to-z$, and the time--dependent ansatz
\bea\label{7D metric}
g_{MN}=\diag(-n^2,a^2\,\z_{ij},b^2\,\xi_{ab},f^2)  \non
\ena
All the warp factors are functions of $t$ and $z$, $\z_{ij}$ and $\x_{ab}$ are 3D and 2D maximally symmetric metrics with spatial curvatures $k$ and $\k$ respectively. The general action including
matter in the bulk is $S_{RS}=S_{EH}+S_{m_B}+S_{V}+S_{m_b}$ ($S_{EH}$ stands for the bulk Einstein--Hilbert action with cosmological constant $\L_7$, $V$ is the tension of the brane, $S_{m_B}$ and $S_{m_b}$
denote the matter actions in the bulk and on the brane respectively). The induced 6D metric on the brane is given by $\g_{\m\n}=\diag(-1,a^2\z_{ij},b^2\xi_{ab})$. Moreover, the RS fine--tuning is expressed as
$3V^2=40M^5\L_7$.

The RS holographic dual can be derived following \cite{Hawking:2000kj}. On--shell gravity in a $(d+1)$--dimensional space--time with boundary is regularized imposing a cut--off at the boundary. A finite
number of counterterms are needed in order for the action not to diverge in the limit in which the cut--off goes to zero (i.e. when the full space--time is recovered). The result of this holographic
renormalization procedure  \cite{Bianchi:2001kw} is a renormalized gravity action, where only the counterterm of order $d/2$ in the curvature is cut--off dependent. 
\vspace{-0.15cm}\sectioncount{Brane/bulk correspondence}\label{correspondence}
Renormalized gravity in 7D is given by $S_{gr}=S_{EH}+S_{GH}-S_{count}$ ($GH$ stands for Gibbons--Hawking, the boundary counterterm actions $S_{count}=\sum_0^3S_i$ are of order $i$ in the curvature).  AdS/CFT
correspondence \cite{Maldacena:1997re} states that the low energy limit of gravity in a $(d+1)$D background with $N$ $d$--branes can be holographically described by the large--$N$ CFT living on the branes
$Z_{gr}\left[\phi_i\right]=\exp\{-W_{CFT}\left(\phi_i\right)\}$.  Using this relation, in analogy to \cite{Kiritsis:2005bm}, we infer that the dual of 7D RS $S_{RS}=S_{EH}+S_{GH}+S_{V}+S_{m_b}$ ($S_{GH}$
arises as a boundary term, $\g$ is the 6D metric) is given by $S_{\tilde{RS}}=2W_{CFT}+2S_1+2S_2+2S_3+S_{m_b}$ if we fine--tune $S_{V}=-2S_0$ (no effective cosmological constant on the brane).
$S_1$ gives the 6D $EH$ action with Planck mass $\Mpl=M^5\ell$, where $\ell$ is the length scale of the background (the 4D Planck mass is $M^2_{\scriptscriptstyle{(4)}}=V_{\scriptscriptstyle{(2)}}\Mpl$,
$V_{\scriptscriptstyle{(2)}}\sim b^2$ is the volume of the compact space --- variable in time except for static 2D space). The Einstein equations are derived in \cite{Mazzanti:2007sg}, considering matter as a
perfect fluid with density $\rho$, 3D pressure $p=w\rho$ and 2D pressure $\p=w_\p\rho$. The (non traceless \cite{Bianchi:2001kw}) energy tensor associated to the variation of $2(W_{CFT}+S_2+S_3)$ is
parametrized with density $\s$, 3D and 2D pressures $\s_p$ and $\s_\p$.

Neglecting the contributions from higher orders in the curvature, namely $S_2$ and $S_3$, in the homogeneous limit where 3D and 2D spaces evolve according to the same Hubble parameter $H\equiv\dot a/a$ and
$\p=p$, yields the following Friedmann plus conservation equations on the brane
\bea
H^2+\oneover{10}\frac{\k}{a^2}&=&\oneover{10\Mpl}\left(\rho+\chi\right)\label{H 6D equal linear}\\
\dot\chi+6H\chi&=&0, \quad
\dot\rho+\weq H\rho=0\label{rho 6D equal linear}
\ena
Here $k=0$ and $\weq\equiv5(1+w)$. The mirage density $\chi$ is identified with the solution to the homogeneous equation associated to the equation for $\s$ and, in this approximation, $\chi=\s$.
Eqs.~\refeq{H 6D equal linear}--\refeq{rho 6D equal linear} perfectly match with the bulk view--point results
\bea\label{H 7D equal linear}
H^2+\oneover{10}\frac{\k}{a^2}&=&\frac{V}{200M^{10}}\left(\rho+\chi\right)\\
\dot\chi+6H\chi&=&0, \quad
\dot\rho+\weq H\rho=0\label{rho 7D equal linear}
\ena
if the gravity parameters are related by $\Mpl=20M^{10}/V$ --- which is exactly the RS fine--tuning condition. The mirage density in the bulk description is defined by its differential equation \refeq{rho 7D
equal linear} and comes from the reduction of the first order ODE for $H$ to the Friedmann equation \refeq{H 7D equal linear}, which is truncated to linear order in $\rho$ (quadratic dependence on the matter
density appears in the full treatment, also on the brane side). 
\sectioncount{Bulk evolution with energy exchange}\label{evolution}
The general Friedmann and modified conservation equations in the case of non zero energy exchange --- parametrized as $T=A\rho^\n$ (a 5D/4D analogue is described in \cite{Kiritsis:2002zf}) --- yield a wide
range of possible cosmologies. Interesting features arise in the case of energy influx $T<0$. The number of critical point solutions varies depending on $\n$. As an example, a typical two point solution is
represented in figure below for flat ($\k=0$) and static compact extra dimensions $b\equiv b_0$, where the dimensionless acceleration $\cqq=\g^2q$ ($q\equiv\ddot a/a$) is plotted as a function of the 6D
dimensionless density $\crho=\g^6\rho$ (the 4D effective density is $\crhofo=V_{\scriptscriptstyle{(2)}}\crho$, $\g^4\sim V/M^{10}$). 
\begin{figure}[ht]\label{ph sp static}
\includegraphics[width=0.4\textwidth]{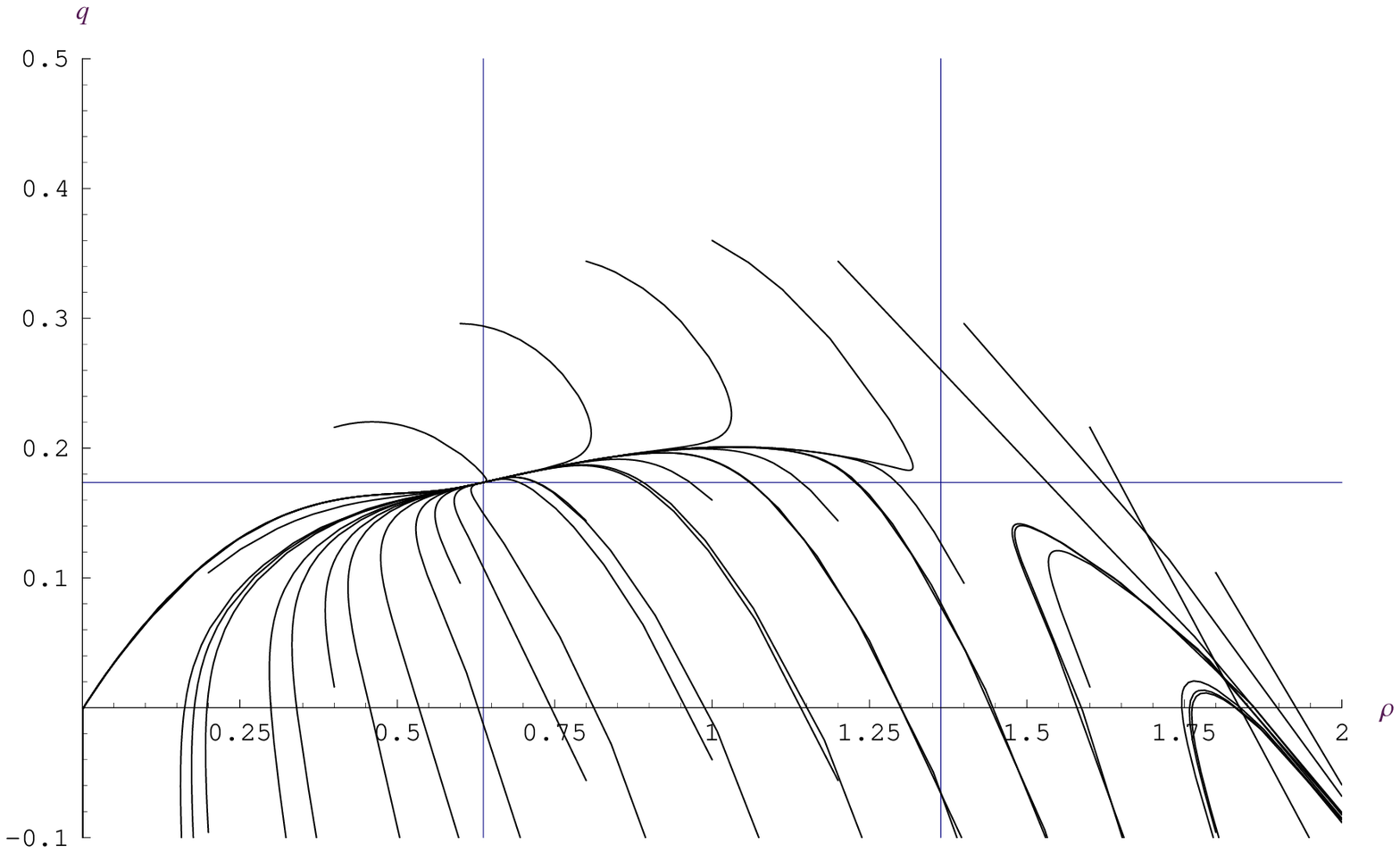}
\end{figure}
Accelerated and decelerated eras can alternate for the trajectories in this phase space. More generally, we can find de Sitter solutions which are stable for a large choice of the parameters in the theory,
thus eventually representing the present acceleration of the universe. Eras with larger acceleration may be identified with primordial inflation. Intermediate phases with negative acceleration can show up, if
we argue that $T$ changes from outflow to influx after reaching a limiting value for $\rho$ at non equilibrium. For energy outflow $T>0$ and $b\equiv b_0$, we find that all trajectories flow to the trivial
critical point $H=\rhofo=0$, eventually passing through an accelerated phase. A more detailed analysis is achieved in \cite{Mazzanti:2007sg}.

\end{document}